\documentclass[epsf,aps,twocolumn,nofootinbib]{revtex4}
\usepackage{amssymb}

\usepackage{graphicx}
\usepackage{amsmath}
\usepackage{epsfig}

\def\ba{\begin{eqnarray}}
\def\ea{\end{eqnarray}}
\def\B{\begin{eqnarray}}
\def\E{\end{eqnarray}}
\def\be{\begin{equation}}
\def\ee{\end{equation}}
\def\C{\mathcal{C}}
\def\K{\mathcal{K}}
\def\H{\mathcal{H}}
\def\A{\mathcal{A}}

\def\dd{\left|\partial d\right|}

\def\nn{\nonumber}
\def\cosech{\mathrm{cosech}}
\def\cosec{\mathrm{cosec}}

\def\x{\mathbf{x}}
\def\d{\mathrm{d}}

\def\mn{{\mu\nu}}
\def\gap{\phantom{(+)}}
\def\OO{\hat{O}}
\def\dt{\delta\tau}

\newcommand{\nc}{\newcommand}
\nc{\taup}[2]{\tau^{(+) #1}_{\phantom{(+)}#2}}
\nc{\taum}[2]{\tau^{(-) #1}_{\phantom{(+)}#2}}
\nc{\taupm}[2]{\tau^{(\pm) #1}_{\phantom{(+)}#2}}
\nc{\tauhp}[2]{\hat{\tau}^{(+) #1}_{\phantom{(+)}#2}}
\nc{\tauhm}[2]{\hat{\tau}^{(-) #1}_{\phantom{(+)}#2}}
\nc{\tauhpm}[2]{\hat{\tau}^{(\pm) #1}_{\phantom{(+)}#2}}
\nc{\Dp}{D^{(+)}}
\nc{\Dm}{D^{(-)}}
\nc{\gp}{g^{(+)}}
\nc{\gpm}{g^{(\pm)}}
\nc{\pmix}{^{(+)\mu}_{\gap\nu}}
\nc{\mmix}{^{(-)\mu}_{\gap\nu}}
\nc{\pmmix}{^{(\pm)\mu}_{\gap\nu}}
\nc{\mix}{^\mu_\nu}
\nc{\nmix}[1]{^{\mu(#1)}_{\nu}}
\nc{\binomial}[2]{\Big(\begin{array}{c}#1\\#2\end{array}\Big)}
\nc{\dm}{\partial_\mu}
\nc{\dn}{\partial_\nu}

\begin{document}

\title{High-energy effective theory for matter on close Randall Sundrum branes }

\author{Claudia de Rham{\footnote{e-mail address:
      C.deRham@damtp.cam.ac.uk}} and Samuel Webster{\footnote{e-mail address:
      S.L.Webster@damtp.cam.ac.uk}}\vspace{5mm}}

\affiliation{Department of Applied
Mathematics and Theoretical Physics\\
University of Cambridge \\
Wilberforce Road, Cambridge CB3 0WA, England}

DAMTP-2005-55

%
\begin{abstract}
Extending the analysis of \cite{deRham:2005xv}, we obtain a formal
expression for the coupling between brane matter and the radion in
a Randall-Sundrum braneworld. This effective theory is correct to all
orders in derivatives of the radion in the limit of small brane
separation, and, in particular, contains no higher than second
derivatives. In the case of cosmological symmetry the theory can be
obtained in closed form and reproduces the five-dimensional
behaviour. Perturbations in the tensor and scalar sectors are then studied. When the branes are
moving, the effective Newtonian constant on the brane is shown to
depend both on the distance between the branes and on their velocity. In
the small distance limit, we compute the exact dependence between the
four-dimensional and the five-dimensional Newtonian constants.\vspace{45pt}
\end{abstract}

\maketitle

\section{Introduction}
Advances in string/M-theory have recently motivated the study of new
cosmological scenarios for which our Universe would be embedded in
compactified extra dimensions where one extra dimension could be
very large relative to the Planck scale. Although the notion of extra dimensions
is not new \cite{Kaluza:1921tu,Klein:1926tv,Antoniadis:1990ew},
 braneworld scenarios offer a new approach for realistic cosmological models.
In some of these models, spacetime
is effectively five-dimensional and gauge and matter fields are
confined to three-branes while gravity and bulk fields propagate in the whole
spacetime \cite{Gibbons:1986wg,Brax:2004xh,Davis:2005au,Langlois:2002bb}.
Playing the role of a toy model, the Randall Sundrum (RS) scenario is of
special interest \cite{Randall:1999ee}. In the RS model, the extra dimension is compactified
on an $S_1 / \mathbb{Z}_2$ orbifold, with two three-branes (or
boundary branes) at the fixed points of the $\mathbb{Z}_2$
symmetry. In the model no bulk fields are present and only gravity
propagates in the bulk which is filled with a negative cosmological constant.
In the low-energy limit, an effective four-dimensional theory can be
 derived on the branes
 \cite{Mendes:2000wu,Khoury:2002jq,Kanno:2002ia,Shiromizu:2002qr}.
However, beyond this limit, braneworlds models differ
remarkably from standard four-dimensional models and have some distinguishing elements
which could either generate cosmological signatures or provide
 alternative scenarios to standard cosmology.
This has been pointed out in
many publications \cite{Langlois:2003yy,Maartens:1999hf,Langlois:2000ns,Copeland:2000hn,
Liddle:2001yq,Sami:2003my,Maartens:2003tw,Koyama:2004ap,
deRham:2004yt,Calcagni:2003sg,Calcagni:2004bh,Papantonopoulos:2004bm,
Kunze:2003vp,Liddle:2003gw,Shiromizu:2002ve} and in particular in \cite{deRham:2005xv}
where the characteristic features of the model are pointed out in the limit
that two such three-branes are close to each other. In particular the
effective four-dimensional theory
was derived in the
limit when the distance between the branes is much smaller than the
length scale characteristic for the five-dimensional Anti-de Sitter (AdS)
bulk. The effective four-dimensional Einstein
equations are affected by the braneworld nature of the model and
 new terms in the Einstein equations contain arbitrary
powers of the first derivative of the brane distance. In \cite{deRham:2005xv} the main results have
been obtained for the simple case where no matter is present on the branes.

The main purpose of this work is to extend this analysis  and to
derive
an effective theory in the presence
of matter on the branes. At high energy, matter couples to gravity in a
different way to what is usually expected in a standard four-dimensional
scenario. In particular gravity couples quadratically to the
stress-energy tensor of matter fields on the brane as well as to the
electric part of the five-dimensional Weyl tensor, which encodes
information about the bulk geometry
\cite{Shiromizu:1999wj,Binetruy:1999hy,Flanagan:1999cu,Mukohyama:1999qx}. Consequently we
expect the covariant theory in presence of matter to be genuinely
different than normal four-dimensional gravity and to bring some
interesting insights on the way matter might have coupled to gravity
at the beginning of a hypothetical braneworld Universe.

Our analysis relies strongly on the assumption that the brane separation
is small, so that the results would only be valid just after or just
before a collision. However, it is precisely this regime that is of
great
importance if one is to interpret the Big Bang as a brane
collision
\cite{Khoury:2001wf,Khoury:2001bz,Webster:2004sp,Gibbons:2005rt,Jones:2002cv}
or as a collision of bubbles \cite{Gen:2001bx}.
In particular, we may point out \cite{Blanco-Pillado:2003hq}
where it is shown that bubbles collision could lead to a Big Crunch.
The authors show that, close to the collision, the bubbles could be treated
as branes. Their collision would lead to a situation where the
branes are sticking together, creating a spatially-flat expanding
Universe, where inflation could take place.
In that model, the collision will be well defined and not lead to any five-dimensional
singularities.

In order to study the presence of matter on the branes in a model
analogue to RS, we first derive, from the five-dimensional theory, the exact Friedmann equations on the
branes for the background. This is
done in section \ref{5d}, in the limit where the branes are close
together, i.e. either just
before or just after a brane collision.  We then give in section
\ref{sec eff theory} an overview of
the effective four-dimensional theory in the limit of small brane
separation as presented in \cite{deRham:2005xv} and show how the
theory can be formally extended in order to accommodate the presence of
matter on the boundary branes. We then check that this theory gives a
result that agrees perfectly with the five-dimensional solution for
the background. Having checked the consistency of this effective
theory for the background, we use it in order to study the effect of
matter perturbations about an empty background (i.e. a `stiff source' approximation) in section
\ref{sec perturbations}. For this we
consider the branes to be empty for the background and introduce
matter on the positive-tension brane only at the perturbed level. We then
study with more detail the propagation of tensor and scalar
perturbations. Although the five-dimensional nature of the theory does
not affect the way tensor and scalar perturbations propagate in a
given background, the coupling to matter is indeed affected. In particular, we show that
the effective four-dimensional Newtonian constant depends both on the
distance between the branes and their rate of separation. We then extend the analysis in order to
have a better insight of what might happen when the small brane
separation condition is relaxed. The implications of our results are
discussed in section \ref{sec conclusion}.
Finally, in appendix \ref{appendix scalar perturb}, we present the technical
details for the study of scalar perturbations within this close-brane
effective theory.
%
%
\section{Five-dimensional background behaviour}
\label{5d}
We consider a Randall-Sundrum two brane model allowing the presence of
general stress-energies on each brane. Specifically, we assume an
action of the form
\B
\label{5daction}
S&=&\int \d^5 x \sqrt{-g}
\left(\frac{1}{2\kappa^2}R-\Lambda\right)\\
&&+\sum_{i=\pm} \int_{\mathcal{M}^i}\d^4 x\sqrt{-g^i}\left(-\lambda_i +
\mathcal{L}_i\right),\nn
\E
where the two four-dimensional integrals run over the positive- and
negative-tension branes $\mathcal{M}^\pm$ respectively and $g^\pm_\mn$ are the induced metrics. We
assume a $\mathbb{Z}_2$ symmetry across the branes.

The five-dimensional bulk is filled with a negative cosmological
constant
$\Lambda=-6/ \kappa^2 L^2$, where $L$ is the associated AdS radius and
$\kappa^2$ the five-dimensional Newtonian constant.
The tensions $\lambda_\pm$ are, without loss
of generality, assumed to take their standard fine-tuned values
\be
\lambda_\pm=\pm\frac{6}{\kappa^2 L},
\ee
with any deviations from these absorbed into the matter Lagrangians
$\mathcal{L}_\pm$. The resulting four-dimensional stress-energy tensors on the
brane can be written as
\[
T^\pm_\mn=\mp\frac{6}{\kappa^2 L}g^\pm_\mn+\tau^{(\pm)}_\mn.
\]
In this paper we use the index conventions that Greek indices run from
0 to 3 and Latin from 1 to 3, referring to the Friedmann-Robertson-Walker (FRW) coordinate systems
defined below in (\ref{coords}).

The point of this section is to extract as much information as
possible about the dynamics of the system in the case of cosmological
symmetry in order to obtain a result against which the effective
theory can be checked. Therefore, we both assume the bulk and the
brane stress-energies $\tau^\pm_\mn$ to have the
required symmetries. Generalising the analysis of
\cite{deRham:2005xv},
we work again in the stationary
Birkhoff frame:
\B
\d s^2&=&\d Y^2-n^2(Y)\d T^2+a^2(Y)\d\x^2 \nn\\
\label{a and n}
\text{with\ \ \ } a^2(Y)&=&e^{-2 Y/L}+\frac{\mathcal{C}}{4}\ e^{2 Y/L}\\
n^2(Y)&=&L^2a'(Y)^2
=a^2-\frac{\C}{a^2},\nn
\E
with flat spatial geometry for simplicity. The trajectories of the
branes are $Y=Y_\pm(T)$ giving the induced line elements
\B
\d s_{\pm}^2&=&-( n^2_\pm-\dot{{Y}}_{\pm}^2) \d T^2+
a_{\pm}^2 \d \x^2
\nn\\
\label{coords}
&\equiv& -\d t_{\pm}^2+a_\pm^2(t_\pm)\d\x^2,
\E
where $a_\pm(T)=a\left(Y=Y_\pm(T)\right)$ and similarly for
$n_\pm(T)$. The velocities of the branes are completely
prescribed by the Isra\"el junction conditions \cite{Israel:1966rt}:
\B
\label{velocities}
\frac{\d Y_\pm}{\d
  T}^2&=&n_\pm^2\left(1-\frac{n_\pm^2}{a_\pm^2}F_\pm\left(\rho_\pm\right)\right)\\
\label{F}
F_\pm\left(\rho_\pm\right)&=&\left(1\pm \frac{\kappa^2
  L}{6}\rho_\pm\right)^{-2},
\E
where $\rho_\pm$ are the brane energy densities $-\taupm{0}{0}$.
The Hubble parameter on each brane then
follows as
\B
\label{hubble}
H_\pm^2&=&\frac{1}{L^2}\left(\frac{1}{F_\pm\left(\rho_\pm\right)}
-\frac{n_\pm^2}{a_\pm^2}\right)\\
\label{5dquadterms}
&=&\frac{\C}{L^2
    a_\pm^4}\pm\frac{\kappa^2}{3L}\rho_\pm+\frac{\kappa^4}{36}\rho_\pm^2.
\E
As in \cite{deRham:2005xv} we now consider the limit of small brane
    separation by replacing $n_\pm$ and $a_\pm$ with their values
    $n_0$ and $a_0$ at the collision (equivalent to taking the leading
    order in $d/L$ where $d$ is related to the radion, as defined
    below).

 To this level of approximation the brane
    position are then given by
\B
\label{Y(T)}
Y_\pm(T)&\sim& Y_0\mp v_\pm\left(T-T_0\right)\\
\label{vpm}
v_\pm&=&n_0\sqrt{1-\frac{n_0^2}{a_0^2}F_\pm\left(\rho_\pm(T=T_0)\right)},
\E
where here and subsequently we take $\sim$ to denote the leading order
    in $d/L$, and we have chosen to consider the motion of the branes
    immediately after a collision at $T=T_0$ and $Y=Y_0$ when the
    branes are moving apart. Note firstly that the branes will in
    general be moving with different velocities. Secondly, the limit
    of large energy density $\rho_\pm\rightarrow \infty$
    corresponds to $v_\pm\rightarrow n_0$, i.e. the limit of null
    brane velocity.

The transformation
\B
T-T_0&=&\frac{t}{n_0}\cosh{\alpha(y)}\nn\\
\label{xfmn}
Y-Y_0&=&t\ \sinh{\alpha(y)}\\
\alpha(y)&=&(y-1)\tanh ^{-1}\left(\frac{v_+}{n_0}\right) +
y\tanh^{-1} \left(\frac{v_-}{n_0}\right),\nn
\E
brings the  brane loci (\ref{Y(T)}) to the fixed
positions $y=0,1$, with line element
\B
\label{bulklineelement}
\d s^2&\sim&d(t)^2 \d y^2-\d t^2+a(y,t)^2 \d\x^2\\
d(t)&=&t\left(\tanh^{-1}\left(\frac{v_+}{n_0}\right)
+\tanh^{-1}\left(\frac{v_-}{n_0}\right)\right).\nn
\E
Note as a consistency check that the global coordinate $t$ coincides
for $y=0,1$ with the proper times
$t_\pm$ on the two branes (in the small $d$ limit) as defined in (\ref{coords}), e.g.
\[
\left.\frac{\d t}{\d T}\right|_{y=0}=\sqrt{n_0^2-v_+^2}\sim \frac{\d
  t_+}{\d T}.
\]
 A generalisation of this metric to
 \be
\label{general metric 1}
\d s^2=A(x,y)^2 \d y^2 + q_\mn(x,y)\d x^\mu \d x^\nu,
\ee
for branes
fixed at $y=0,1$ is the starting point for the derivation of the
effective theory in the next section.
There, the proper distance between the
two branes is measured along a trajectory of constant $x^\mu$, i.e. it
is taken to be
\be
\label{def of d}
d(x)=\int_0^1 \d y\ A(x,y)^2.
\ee
In particular, if we choose a
specific gauge for which $A(x,y)$ is independent of $y$, the metric
(\ref{general metric 1}) is simply
\be
\label{general metric}
\d s^2=d(x)^2 \d y^2 + q_\mn(x,y)\d x^\mu \d x^\nu.
\ee
As discussed in \cite{deRham:2005xv}, it is unclear whether such a
gauge may be fixed in general, but it can be shown that the
resulting effective theory is not sensitive to the $y$ dependence of
$A$.

From (\ref{hubble}) and (\ref{vpm}) it can then be shown that
the
Hubble parameter at the time of collision is related to the rate of expansion
of the fifth dimension with respect to proper time $t$ by
\B
\label{5dresult}
H_+(0)&=&\frac{1}{L}\tanh \frac{\dot{d}(0)}{2}\\&& +
\frac{\kappa^2}{6}\left(\rho_+(0) \coth\ \dot{d}(0) + \rho_-(0) \cosech\
\dot{d}(0)\right).\nn
\E

\begin{figure}
\begin{center}
\includegraphics[width=9cm]{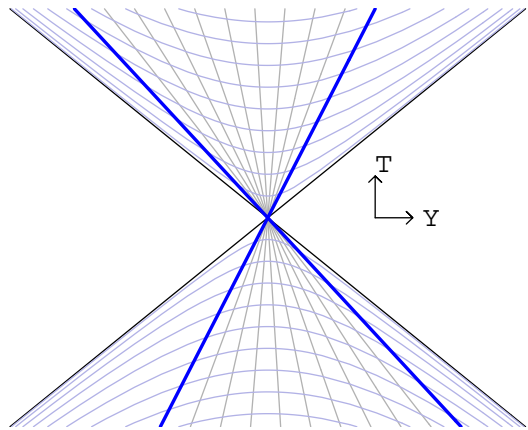}
\caption{Comparing coordinate systems: In the near-brane limit, the
  two branes move along the lines $Y=Y_\pm(T)$ (thick lines). Lines of
  constant $y$ (straight) and $t$ (curved) are shown; $d(t)$ is the proper distance along
  a line of constant $t$ between the two branes. Note that, if
  $v_+\neq v_-$, the values of $T$ at either end of this line will be
  different. The coordinate system $(y,t)$ is only defined inside the
  lines $Y-Y_0=\pm n_0(T-T_0)$ as shown}
\label{YTyt}
\end{center}
\end{figure}

Note that this is in
general not the same as other definitions of the radion, more common ones being the
distance along integral curves of the normal (lines of constant $x$
are not in general geodesics in this metric, see Fig. 1) or, different
again,
$Y_-(T)-Y_+(T)$. When the effective
theory is defined from a moduli space approximation, the radion often
enters via a ratio of the conformal factors on the branes, but this is
not meaningful apart from in the small-velocity limit. It is however
of
note that all these definitions are proportional in the
special case of cosmological symmetry and small brane separation.
%
%
\section{Close branes effective theory description}
\label{sec eff theory}
\subsection{Formalism}
We work in a frame where the branes are assumed to be exactly static
at $y=0,1$ with metric (\ref{general metric}) in order to simplify the implementation of the Isra\"el
junction conditions, which would otherwise be difficult. From the Gauss equations, the Einstein tensor on a
$y=\text{const}$ hypersurface is given by \cite{Binetruy:1999ut,Shiromizu:1999wj}:
\ba
G^\mu_\nu(y)&=&\frac{3}{L^2}\delta^\mu_\nu+K
K^\mu_\nu-K^\mu_\alpha K^\alpha_\nu \label{GuvGC}\\
&&-\frac 1 2 \delta^\mu_\nu
\left(K^2-K_\alpha^\beta K^\alpha_\beta \right) -E^\mu_\nu. \notag
\ea
The unknown quantity in (\ref{GuvGC}) is the electric part of the
 projected Weyl tensor $E^\mu_\nu$ which is traceless, enabling us to write the Ricci scalar purely in terms of the extrinsic curvature:
\ba
R=-\frac{12}{L^2}-K^2+K_\alpha^\beta K^\alpha_\beta . \label{R}
\ea
The Weyl tensor $E^\mu_\nu$ can be expressed in terms of
 more recognisable quantities as
\ba
E^{\mu}_{\nu}=-\frac{1}{d} \frac{\partial}{\partial y}K^{\mu}_{\nu}-\frac{1}{d}D^{\mu}D_{\nu}
d-K^{\mu}_{\alpha}K^{\alpha}_{\nu}+\frac{1}{L^2}\delta^{\mu}_{\nu},
\label{Euv1}
\ea
where $D_\mu$ is the covariant derivative with respect to $q_{\mu
\nu}(y)$, implying from (\ref{GuvGC}) that:
\ba
G^\mu_\nu=\frac{1}{d}\hspace{-10pt}&&\Big[
D^{\mu}D_{\nu}d+\frac{\partial}{\partial y}K^{\mu}_{\nu}\label{Guv2}\\
&& +\frac{2 d}{L^2}\delta^\mu_\nu+d K K^\mu_\nu
-\frac d 2 \delta^\mu_\nu
\left(K^2-K_\alpha^\beta K^\alpha_\beta \right)
\Big],\notag
\ea
where the second line is of higher order in the small distance limit $d \ll
L$.

 In order to find the derivative of the extrinsic curvature
 $\frac{\partial}{\partial y}K^{\mu}_{\nu}$ on the brane,
we consider the Taylor
expansion:
\ba
K^{\mu}_{\nu}(y=1)=\sum_{n = 0}^{\infty} \frac{1}{n!}
\left. \partial _{y}^{(n)}
K^{\mu}_{\nu} \right|_{y=0}. \label{taylor2}
\ea
We expand the $n^{\text{th}}$ derivative of the extrinsic curvature in
powers of $d/L$, keeping only the leading term:
\ba
\left.\frac{\partial^n}{\partial y^n}K^{\mu}_{\nu}\right|_{y=0}
\equiv K^{\mu \, (n)}_{\nu}=\K^{\mu \, (n)}_{\nu} \left(1+\mathcal{O}\left(\frac{d}{L}\right) \right)\notag
\ea
and, as shown in Appendix \ref{derivation}, one can obtain the recurrence relation
\ba
\K^{\mu \, (n)}_{\nu}=\hat O \K^{\mu \, (n-2)}_{\nu},
\ea
where the operator $\hat O$ is defined by
\ba
\hat O Z^{\mu}_{\nu}= \left[
d^{,\mu}Z^{\alpha}_{\ \nu}
+d_{,\nu} Z^{\alpha \mu}-
d^{,\alpha} Z^{\mu}_{\ \nu}
\right]d_{,\alpha},
\ea
giving
\ba
\hspace{-10pt}\sum_{n=0}^{\infty} \frac{1}{n!} \K^{\mu \, (n)}_{\nu}
&=&
\sum_{n=0}^{\infty} \frac{1}{(2n+1)!}\hat O^n  \K^{\mu \, (1)}_{\nu}
\notag
+\frac{1}{(2n)!}\hat O^n  \K^{\mu \, (0)}_{\nu}
\notag  \\
&=& \frac{\sinh \sqrt{\hat O}}{\sqrt{\hat O}}
 \ \K^{\mu \, (1)}_{\nu}
+ \cosh \sqrt{\hat O} \ \K^{\mu \, (0)}_{\nu}.
\ea
We then have a formal expression for the first derivative of the
extrinsic curvature in terms of the radion
and stress-energy:
\ba
\left.\frac{\partial}{\partial y}K^{\mu}_{\nu}\right|_{y=0}
\sim-\frac{\sqrt{\hat O}}{\sinh \sqrt{\hat O}}
\Big[
\cosh \sqrt{\hat O}\, K\pmix
-K\mmix \label{K'+}
\Big],
\ea
where $K\pmmix=K^{\mu}_{\nu}(y=0,1)$. It is straightforward then to obtain the corresponding result for the negative-tension brane:
\B
\left.\frac{\partial}{\partial y}K^{\mu}_{\nu}\right|_{y=1}&=&\left.\frac{\partial}{\partial y}K^\mu_\nu\right|_{y=0}+\sum_{n=2}^\infty \frac{K^{\mu(n)}_{\nu}}{\left(n-1\right)!}\nn\\
&\sim&\left.\frac{\partial}{\partial y}K^\mu_\nu\right|_{y=0}+\left(\cosh\sqrt{\hat O}-1\right)\left.\frac{\partial}{\partial y}K^\mu_\nu\right|_{y=0}\nn\\&&\quad+\sinh\sqrt{\hat O} K\pmix\nn\\
&\sim&\frac{\sqrt{\hat O}}{\sinh \sqrt{\hat O}}
\Big[
\cosh \sqrt{\hat O}\, K\mmix
-K\pmix \label{K'-}
\Big].
\E
%
\subsection{Einstein equations on the branes}
The next step is to use the Isra\"el junction conditons to rewrite the extrinsic
curvatures of the two branes in
terms of the stress-energy tensors and the tensions:
\B
\label{kmunu}
K\pmix&=&-\frac{1}{L}\delta^\mu_\nu-\frac{\kappa^2}{2}\tauhp{\mu}{\nu}\\
\label{kmunu2}
K\mmix&=&-\frac{1}{L}\delta^\mu_\nu+\frac{\kappa^2}{2}\tauhm{\mu}{\nu}\\
\label{tauhat}
\tauhpm{\mu}{\nu}&\equiv&\taupm{\mu}{\nu}-\frac{1}{3}\taupm{}{}
\delta^\mu_\nu .
\E
This gives us both the value of the Ricci scalar on the branes (\ref{R})
\ba
R^{(\pm)}=\mp \frac{\kappa^2}{L}\tau^{(\pm)}+\frac{\kappa^4}{4}
\left(\hat{\tau}^{(\pm)\,2}-\hat{\tau}^{(\pm)\, \alpha}_{\phantom{(\pm)} \, \beta}
\hat{\tau}^{(\pm)\, \beta}_{\phantom{(\pm)} \, \alpha} \right)\label{R2}
\ea
and, substituting (\ref{K'+}) (or (\ref{K'-})) into (\ref{Guv2}),
the effective Einstein equations:
\B
G\pmmix&\sim&\frac{1}{d}D^{(\pm)\mu}D^{(\pm)}_{\gap \nu} d
\pm\frac{1}{Ld}\dd\tanh\frac\dd 2\delta^\mu_\nu
\label{effectiveeinstein} \\
&&\pm\frac{1}{Ld}\frac{1}{\dd}\left(\tanh\frac\dd
2+\tan\frac\dd 2\right)\partial^\mu d\, \partial_\nu d\nn\\
&&+\frac{\kappa^2}{2d}\A^{(\pm)\mu}_{\gap\nu}
 \nn\\ &&
+\left[\pm \frac{2\kappa^2}{L}\left(\taupm{\mu}{\nu}
-\frac{\tau^{(\pm)}}{6}\delta^\mu_\nu\right)+\kappa^4
\Pi^{(\pm)\mu}_{\gap\nu}\right],\nn
\E
where
\be
\label{coupling}
\A^{(\pm)\mu}_{\gap\nu}=\sqrt{\OO}\left(\coth
\sqrt{\OO}\tauhpm{\mu}{\nu}
+\cosech\sqrt{\OO}\hat{\tau}^{(\mp)\mu}_{\gap\nu}\right),
\ee
and
\be
\Pi^{(\pm)\mu}_{\gap\nu}=\frac{1}{8}\taupm\alpha\beta\taupm\beta\alpha\delta^\mu_\nu
-\frac{1}{12}\tau^{(\pm)}\taupm\mu\nu-\frac{1}{72}\tau^{(\pm)\, 2}\delta^\mu_\nu.
\ee From the tracelessness of $E^{(\pm)}_\mn$ we obtain two equivalent equations of
motion for the radion,
\B
\label{kleingordon}
\hspace{-13pt}\Box^{(\pm)} d&\sim&\mp\frac{\dd}{L}
\left(3\tanh\frac{\dd}{2}-\tan\frac{\dd}{2}\right)
-\frac{\kappa^2}{2}\A^{(\pm)} \\
&&
+ \left[\pm\frac{\kappa^2
  d}{3L}\tau^{(\pm)}- \frac{\kappa^4
    d}{4}\tauhpm{\mu}{\nu}\tauhpm{\nu}{\mu}\right].\nn
\E
The terms in square brackets in (\ref{effectiveeinstein}) and
(\ref{kleingordon}) will turn out to be of higher order as
$d\rightarrow 0$ and so should not strictly be included. However, for
exact cosmological symmetry,
they are the only
higher order terms and we shall keep them for the time being. Later on they shall be discarded.

The conservation of the stress-energy tensor on both branes
follows directly from the Codacci equation \cite{Binetruy:1999ut,Shiromizu:1999wj}:
\ba
D_\mu K^\mu_\nu-D_\nu K=0,
\ea
which, evaluated on the branes implies
\ba
\label{conservation}
D^{(+)}_\mu \taup{\mu}{\nu}=D^{(-)}_\mu\taum{\mu}{\nu}=0.
\ea
%
%
\subsection{Low-energy limit}
As a first consistency check of this close-brane theory, we consider
its low-energy limit and compare it with the effective
four-dimensional low-energy theory
\cite{Mendes:2000wu,Khoury:2002jq,Kanno:2002ia,Shiromizu:2002qr}
for small brane separation. In that common limit, both theories should agree.

In the low-energy limit, the magnitude of the stress-energy tensor on
the brane is small compared to the brane tension. Any
quadratic term $\kappa^4 \tau^{(\pm)\, 2}$ is negligible compared to
$\frac{\kappa^2}{L}\tau^{(\pm)}$, so that $\Pi^{(\pm)
  \mu}_{\phantom{(\pm)}\nu}$
may be dropped in (\ref{effectiveeinstein}) and,
from (\ref{R2}), the Ricci tensor on the brane is:
\ba
R^{(\pm)}=\mp \frac{\kappa^2}{L}\tau^{(\pm)}.
\ea
Furthermore in the low-energy limit, the branes are moving slowly,
 $\dd \ll 1$, to linear order in $\dd$, we have:
\ba
\A^{(\pm)\mu}_{\gap\nu}=
\hat{\tau}^{(+)\mu}_{\gap \nu}
+\hat{\tau}^{(-)\mu}_{\gap\nu}.
\ea
The effective Einstein equation on the brane at low energy
is therefore
\ba
G^{(\pm)\mu}_{\gap\nu}&\sim&\frac{1}{d}D^{(\pm)\mu}D^{(\pm)}_\nu d
+\frac{\kappa^2}{2d}\left(\hat{\tau}^{(+)\mu}_{\gap \nu}
+\hat{\tau}^{(-)\mu}_{\gap\nu}\right)\label{Guvlow}\\
&&\pm\frac{1}{Ld}\left(\partial^\mu d\,  \partial_\nu d
-\frac{1}{2}\left(\partial d\right)^2\delta^\mu_\nu\right),\nn
\ea
with the equation of motion for $d$:
\ba
\Box^{(\pm)} d=\pm \frac{1}{L}\left(\partial d\right)^2
+\frac{\kappa^2}{6}\left(\tau^{(+)}+\tau^{(-)}\right).
\ea
We can therefore write (\ref{Guvlow}) in the more common form:
\ba
G^{(\pm)\mu}_{\gap\nu}&\sim&\frac{\kappa^2}{2d}
\left(\tau^{(+)\mu}_{\gap\nu}+\tau^{(-)\mu}_{\gap\nu}\right)\\
&&
+\frac{1}{d}\left(D^{(\pm)\mu}D^{(\pm)}_\nu
d-\Box^{(\pm)} d\right)\nn \\
&&\pm\frac{1}{L d}\left(\partial^\mu d\, \partial_\nu d
+\frac{1}{2}\left(\partial d\right)^2\delta^\mu_\nu \right),\nn
\ea
which is precisely the result we get from the effective low-energy
theory in the close brane limit
\cite{Mendes:2000wu,Khoury:2002jq,Kanno:2002ia,Shiromizu:2002qr,deRham:2004yt}.
%
%
\section{Cosmological Symmetry}
In the most general case, the coupling of the radion to matter on the
branes given by (\ref{coupling}) is intractable. However, we are
concerned here with the case of cosmological symmetry as a check on
the validity of the theory. We take (\ref{bulklineelement}) as our metric and notice
that
\B
\OO \left(\begin{array}{cc}
A&0 \\
0& B\,\delta^i_j\\
\end{array}\right)
=\dot{d}^2
\left(
\begin{array}{rc}
-A&0 \\
0& B\,\delta^i_j\\
\end{array}
\right).
\E
We can then obtain the coupling tensors (\ref{coupling}) in
closed form:
\B
\A^{(\pm)0}_{\gap 0}&=&-\dot{d}\cot\dot{d}\left(\frac{2}{3}\rho_\pm+p_\pm\right)\\
&&-\dot{d}\ \cosec\ \dot{d} \left(\frac{2}{3}\rho_\mp+p_\mp\right)\nn\\
\A^{(\pm)i}_{\gap j}&=&\frac{\dot{d}}{3}
\left(\rho_\pm\coth\ \dot{d}+\rho_\mp\cosech\
\dot{d}\right)\delta^i_j.
\E
The resulting equations of motion follow from
(\ref{effectiveeinstein}),(\ref{kleingordon}) and
(\ref{conservation}):
\B
\label{eff 1}
H_\pm^2&\sim&\frac{1}{d}\left[-\dot{d}H_\pm\pm\frac{\dot{d}}{L}\tanh\frac{\dot{d}}{2}
+\frac{\kappa^2}{6}\A^{(\pm)i}_{\gap
    i}\right]\\
&&\pm\frac{2\kappa^2}{3L}\rho_\pm+\frac{1}{18}\kappa^4\rho_\pm^2\nn\E
\B
\label{eff 2}
\dot{H}_\pm+2H_\pm^2&\sim&\pm\frac{\kappa^2}{6L}\left(\rho_\pm -
3p_\pm\right)-\frac{\kappa^4}{36}\rho_\pm\left(\rho_\pm + 3p_\pm\right)\\
\label{eff 3}
\ddot{d}+3H_+\dot{d}&\sim&\frac{\dot{d}}{L}\left(3\tanh\frac{\dot{d}}{2}-\tan\frac{\dot{d}}{2}\right)\\
&&+\frac{\kappa^2
  d}{L}\left(\frac{\rho_+}{3}-p_+\right)
+\frac{\kappa^2}{2}\left(\A^{(+)0}_{\gap 0}+\A^{(+)i}_{\gap i}\right)\nn\\
&&+\frac{\kappa^4 d}{4}\left(\frac{7}{9}\rho_+^2+p_+^2+\frac{4}{3}\rho_+ p_+\right)\nn\\
\label{cons}
\dot{\rho}_\pm&=&-3H_\pm\left(\rho_\pm+p_\pm\right).
\E
Equations (\ref{cons}) and (\ref{eff 2}) together imply
\be
\label{5dquadterms2}
H_\pm^2\sim\frac{\tilde\C}{L
a_\pm^4}\pm\frac{\kappa^2}{3L}\rho_\pm+\frac{\kappa^4}{36}\rho_\pm^2,
\ee
where $\tilde\C$ is an integration constant which, at this order, can
be identified with the bulk parameter $\C$ via (\ref{hubble}). The
system is now manifestly finite as $d\rightarrow 0$. Note that, apart
from the presence of quadratic terms, (\ref{eff 2}) is the same result
as that obtained from the moduli space approximation and is, in fact,
exact ($d$ decouples as a consequence of the simplicity of the Weyl
tensor for an AdS-Schwarzschild bulk). However, the additional
information from (\ref{eff 3}) gives additional information not
obtainable from a low-energy effective theory.
Since $H_+$ takes a finite value at the collision, the coefficient of
$d^{-1}$ in (\ref{eff 1}) must vanish at $d=0$; this implies then that
\B
\label{confirmation}
H_+(0)&=&\frac{1}{L}\tanh \frac{\dot{d}(0)}{2}\\&& +
\frac{\kappa^2}{6}\left(\rho_+(0) \coth\ \dot{d}(0) + \rho_-(0) \cosech\
\dot{d}(0)\right),\nn
\E
in agreement with the exact result (\ref{5dresult}).
%
%
\section{Effective theory for perturbations}
\label{sec perturbations}
More interesting is the study of cosmological perturbations, for which
a relatively straightforward solution of the above system is also
available. We shall give a few examples here and point out some
interesting features. Throughout we work only with the
positive-tension brane, assuming the negative-tension brane to be
empty, and drop the $\pm$ signs. Also, we shall assume for simplicity that matter on
the positive-tension brane is only introduced at the perturbed level,
i.e. the background solution $a(t),d(t)$ is that obtained from
(\ref{eff 2}) and (\ref{eff 3}) in the absence of matter. We therefore
have
\[
\begin{array}{cclcl}
\tau^\mu_\nu(\x,t)&=& 0 &+& \dt^\mu_\nu\nn\\
d(\x,t)&=&d(t)&+&\delta d(\x,t)\nn\\
g_\mn(\x,t)&=&\bar{g}_\mn&+&\delta g_\mn,
\end{array}
\]
where $\bar{g}_\mn$ is the usual flat FRW metric with scale factor
$a(t)$. However, in the following we will set $\delta d=0$, either
because we are considering tensor perturbations only or because we
choose to work in such a gauge. Hence we shall assume that $d$ takes its
background value.
%
%
\subsection{Tensor Perturbations}
As the simplest starting point we consider perturbations using the
above formalism and we choose to work in conformal time. We take the metric to be
\be
\label{tensor metric}
\d s^2=a(\eta)^2\left(-\d\eta^2+\left(\delta_{ij}+h_{ij}\right)\d x^i
\d x^j\right),
\ee
with the usual transverse traceless conditions
\[
h^i_i=h^i_{j,i}=0
\]
on the perturbation, spatial indices being raised by $\delta^{ij}$. The
resulting Ricci tensor perturbation is then
\B
\delta R_{00}&=&\delta R_{0i}=0,\nn\\
\label{deltartensor}
\delta R_{ij}&=&\left(\H'+2\H^2\right)h_{ij}+\H
h'_{ij}-\frac{1}{2}\bar{\Box}h_{ij},
\E
where $\H=a'/a=aH$, primes denote differentiation with respect to
conformal time $\eta$ and $\bar{\Box}=-\partial_\eta^2+\partial_i
\partial^i$ is the Minkowski space wave operator. We assume that these
gravity waves are sourced by tensor matter at the perturbative level,
i.e.
\[
\dt^\mu_\nu=\delta\hat{\tau}^\mu_\nu=
\left(
\begin{array}{cc}
0&0\\
0&\tau^i_j
\end{array}
\right),\qquad
\tau^i_i=0.
\]
The perturbed Klein Gordon equation for tensor matter just reduces to
  $\delta\left(\Box d\right)\sim 0$, so it is consistent
 to set the scalar perturbation $\delta d$ to zero,
  i.e. to study purely tensor fluctuations. In this case, the equation of motion for the
perturbations follows from (\ref{effectiveeinstein}):
\B
\label{deltaR}
\delta R_{ij}&\sim&\frac{1}{d}\delta\left(D_i D_j\,
d\right)+\frac{\dot{d}}{Ld}\tanh\frac{\dot{d}}{2}\ a^2 h_{ij}\\
&&+\frac{\kappa^2}{2d}\sqrt{\OO}\coth\sqrt{\OO}\
\delta{\hat{\tau}}_{ij}, \nn
\E
where we have now dropped the sub-dominant matter terms. It is straightforward
  to obtain
\B
\delta\left(D_i D_j\, d\right)&=&-\left(\frac{1}{2}h'_{ij}+\H
  h_{ij}\right)d'\nn\\
\sqrt{\OO}\coth\sqrt{\OO}\,
\delta\hat{\tau}_{ij}&=&\dot{d}\coth\dot{d}\ \tau_{ij}.\nn
\E
Equations (\ref{eff 1}) and (\ref{eff 2}) then imply the relation for the background Hubble parameter
\[
\H'+\H^2=0,\quad
\H^2=-\frac{d'}{d}\H+a^2\frac{\dot{d}}{L}\tanh\frac{\dot{d}}{L}.
\]
Putting this all together we obtain,
to leading order in $d$:
\be
\label{tensors}
\hat{\boxdot}\, h_{ij}\equiv
\left[\bar\Box-\frac{d'}{d}\frac{\partial}{\partial\eta}\right]h_{ij}
=-\kappa^2\frac{\dot{d}}{d}\coth\dot{d}\ \delta\tau_{ij}.
\ee

The same calculation repeated subject to the low-energy approximation, not assuming small $d$, is straightforward.
Since the matter is traceless, the standard equations at low energy
\cite{Mendes:2000wu,Khoury:2002jq,Kanno:2002ia,Shiromizu:2002qr,deRham:2004yt} give
\[
R_{ij}\approx\frac{1-\psi}{L^2\psi}\left(2L D^{(+)}_i \partial_j d
 -g_{ij}\partial d^2+2\partial_i d \partial_j
 d\right)+\frac{\kappa^2}{L\psi}\tau^{(+)}_{ij},
\]
where $\psi=1-\exp\left(-2d/L\right)$. Perturbing this gives
\be
\label{msa tensor perturbations}
\bar\Box
h_{ij}-2\left[\H+\frac{\left(1-\psi\right)}{\psi}\frac{d'}{L}\right]h'_{ij}\approx-\frac{2\kappa^2}{L\psi}\delta\tau_{ij},
\ee
using the equations
\[
\H'+\H^2=0,\quad \H^2+2\frac{\left(1-\psi\right)}{L\psi}\H
d'\approx\frac{d'^2}{L^2}\frac{\left(1-\psi\right)}{\psi},
\]
for the background. The small-$d$ limit of (\ref{msa tensor perturbations}), where $\psi\sim 2d/L$,  is then
\[
\bar\Box
h_{ij}-\frac{d'}{d}h'_{ij}\sim-\frac{\kappa^2}{d}\delta\tau_{ij}.
\]
The operator $\hat{\boxdot}$ defined in (\ref{tensors}) is therefore the same as one would find in the low-energy theory. The difference lies in the source term; in the high-energy theory, the effective four-dimensional Newton
constant on the positive-tension brane is related to the five-dimensional one by
\ba
\kappa^{(+)\, 2}_{4d}=\frac{\dot d }{d}
\coth \dot d\ \kappa^2, \label{k4d}
\ea
whereas the low-energy result has
\ba
\kappa^{(+)\, 2}_{4d} \sim
\frac{\kappa^2}{d_0}. \label{k4deff}
\ea
As is the case in the low-energy effective theory, the coupling to
matter is different for the background as it is for the
perturbations - for the background, the coupling can be identified
from (\ref{5dquadterms}) or (\ref{5dquadterms2}) as $\kappa^2/L$, as opposed to (\ref{k4d}).
When either the branes are stabilised, and $d$ is not
treated as a dynamical variable, $d\sim d_0=\text{const}$, or the
velocity is small $\dot{d}\ll 1$ (which is the case
in the low-energy limit), it is easy to see that (\ref{k4d}) and (\ref{k4deff}) agree.
However, for arbitrary brane velocities, when the radion is not
stabilised, the exact result for small
$d$ is given by (\ref{k4d}). As expected, the effective Newton
constant picks up a dependence on $d$, as it does in the low-energy
theory, but more unexpected, it also contains some degree of freedom:
the brane separation velocity. Whilst
this is not expected to be relevant today, since one would assume the
radion is stabilised in the present Universe, it would be extremely
important near the brane collision. As discussed in section \ref{5d},
$\dot{d}$ would be approximately constant, $\dot{d}\sim v$, leading to
\ba
\kappa^{(+)\, 2}_{4d} \sim \frac{\coth v}{t}\, \kappa^2,
\ea
where the coefficient $\coth v$ could take any value
greater than $1$ depending on the matter content of the branes.
%
%
\subsection{Scalar Perturbations}
\label{section scalar perturb}
We now consider scalar metric perturbations on the brane sourced by a
perfect fluid at the perturbative level (again, the background
geometry is taken to be empty). We choose to work in a gauge where
$\delta d=0$, i.e. to evaluate the perturbations on hypersurfaces of
constant $d$, in which the metric perturbation can be taken as
\B
\label{ds scalar}
\d s^2&=& a(\eta)^2\Big(\left(-1+2\Phi\right)\d\eta^2+4 E_{,i}
\d\eta\d x^i\\
&&\qquad\qquad+\left(1+2\Psi\right)\delta_{ij} \d x^i \d x^j\Big).\nn
\E
The calculations are not nearly so straightforward as for tensors and
have therefore been relegated to Appendix \ref{appendix scalar perturb}. The result is the
following equation of motion for the curvature perturbation
\be
\hat{\boxdot}\, \Psi\equiv\left[\bar{\Box}-\frac{d'}{d}\frac{\partial}{\partial\eta}\right]\Psi
=-a^2 \frac{\kappa^2}{6}\frac{\dot d}{d}\coth \dot d \ \delta
\rho,
\ee
giving rise to the same relation between the four-dimensional Newtonian constant
and the five-dimensional as in (\ref{k4d}).
Here again we may check that, apart from the modification of the
effective Newtonian constant on the brane, the perturbations propagate
in the given background exactly the same way as they would if the
theory were genuinely four-dimensional. This is a very important result
for the propagation of scalar perturbations if they are to generate
the observed large-scale structure.
The five-dimensional nature of the
theory does affect the background behaviour but on this
background the perturbations behave exactly the same way as they would
in the four-dimensional theory.

This result is of course only true
 in the close-brane limit, for which the theory contains no higher
 than second derivatives, only powers of first derivatives. When the
 branes are no longer very close to each other, the theory will become
 higher-dimensional (in particular the theory becomes non-local
 in the one brane limit). The presence of these higher-derivative
 corrections (not expressible as powers of first derivatives) is expected
 to modify the way perturbations propagate in a given
 background, mainly because extra Cauchy data would need to be specified,
 making the perturbations non adiabatic \cite{deRham:2004yt}.
However if we consider a scenario for which the
 large-scale structure is generated just after the brane collision,
 the mechanism for the production of the scalar perturbations will be
 very similar to the standard four-dimensional one.
%
%
\subsection{Relation between the four- and five-dimensional Newtonian constant}
The relation (\ref{k4d}) between $\kappa_{4d}^{(+)\, 2}$ and the five-dimensional
constant $\kappa^2$  is formally only valid for small distance between the
branes. However if we consider the analysis of \cite{deRham:2005xv}, we
may have some insights of what will happen if we had not stopped the
expansion to leading order in $d$. Here, terms of
the form $d\,  \ddot d$ and more generally any term of the form $d^n
d^{(n+1)} $ have been considered as negligible in comparison to
$\dot d$ and therefore only the terms of the form $\dot d ^n $ have been
considered in the expansion. In a more general case, when the branes
are not assumed to be very close to each other, any term of the
form $d^n \, d^{(n+1)}$  should be considered and would
affect the relation between the four-dimensional Newtonian constant
and the five-dimensional one. For moving branes, we therefore expect
the relation between $\kappa_{4d}^{(+)\, 2}$ and $\kappa^2$ to be:
\be
\kappa_{4d}^{(+)\,2}=\frac{\kappa^2}{L}\  \Omega \left(\frac{d}{L},  \dot d,
d\,
 \ddot d, \cdots, d^n\,  d^{(n+1)}\right).
\ee
The relation is therefore a functional of $d$: $\Omega\left[d(t)
  \right]$ has an infinite number of independent degree of freedom.

In the low-energy limit, or when the radion is stabilised,
$d \sim d_0=\text{const}$,  the exact expression of $\Omega$ is \cite{Garriga:1999yh}:
\be
\Omega
\rightarrow \Omega\left[d(t)=d_0\right]=\frac{e^{d_0/L}}{2 \sinh d_0/L}
\ee
For close branes, another limit is now known:
 when $d \ll L$,
\be
\Omega\left[d\ll L \right]= \frac{\dot d}{d}\coth
\dot d.
\ee
But in a general case, $\Omega$ (and therefore $\kappa_{4d}^{(+)\, 2}$)
is expected to be a completely dynamical degree of freedom. For the
present Universe the radion is supposed to be stabilised, but in
early-Universe cosmology, the effective four-dimensional Newton
constant could be very different from its present value. It might
therefore be interesting to understand what the constraints on such
time-variation of the Newtonian constant would be and how it would
constrain the brane velocity \cite{Clifton:2005xr,Barrow:1996kc}, or
whether such a time variation could act as a signature for the
presence of extra dimensions.

\section{Conclusions}
\label{sec conclusion}
In the first part of this work, we derived the exact behaviour of FRW
branes in the presence of matter. The characteristic features
come from the presence of the $\rho^2$ terms in the Friedmann equation
and from the `dark energy' Weyl term. In the limit of close
separation we related the contribution of the Weyl term to
the expansion of the fifth dimension. We then used this result to test
the close-brane effective theory that was first derived in
\cite{deRham:2005xv} but now with matter introduced on the branes. For this
we have shown how matter can be included using a formal sum
of operators acting on the stress-energy tensors for
matter fields on both branes. In the general case the action of this sum of
operators on an arbitrary stress-energy tensor would not be available
in closed form, although one could in principle proceed perturbatively.
When a specific scenario is chosen, however, one can make considerable
analytical progress. Assuming cosmological symmetry, the
action of the operators on the stress-energy tensor is remarkably
simple and the sum can be evaluated analytically. We then compared the
result with the exact five-dimensional result
in the limit of small brane separation. As expected both results agree
perfectly. Furthermore we have checked that, in the low-energy limit, our
close-brane effective theory agrees perfectly with the effective
four-dimensional low-energy theory, giving another consistency check.

We then used this close-brane effective theory in order to understand
the way matter couples to gravity at the perturbed level. In order to
do so, we
considered a scenario in the stiff source approximation for which the
background is supposed to be unaffected by the presence of matter and
considered the production of curvature and tensor perturbation sourced
by the presence of matter fields on the brane. Although the
five-dimensional nature of the theory does affect the background
behaviour, we have shown that for a given background the perturbations
propagate the same way as they would in a standard four-dimensional
theory. This is only true in the limit of small brane separation
 and is not expected to be valid outside this regime. However, since
the large-scale structure of the
Universe might have been produced in a period for which the branes
could have been close together (for instance just after a brane
collision initiating the Big Bang), this regime is of special
interest. The fact that the perturbations behave the same way, for a given
background, as they would in a four-dimensional theory is a remarkable
result for the production of the large scale structure which could be
almost unaffected by the presence of the fifth dimension. On the
other hand, the relation
between the four-dimensional Newtonian constant and the
five-dimensional one is however affected by the expansion of the
fifth-dimension. It has been shown in the literature
\cite{Randall:1999ee,Garriga:1999yh}
 that four-dimensional Newtonian constant was dependant
on the distance between the branes, giving a possible explanation of the
hierarchy problem. In this paper we show that the four-dimensional
Newtonian constant also has some dependence on the brane velocity which we
computed exactly in the small-distance limit, which might be able to
provide an observational signature for the presence of extra
dimensions.
Outside the small $d$ regime, we expect the four-dimensional Newtonian
constant to depend on the five-dimensional one not only through
the brane separation velocity $\dot d$ but also on higher derivatives of the
distance between the branes $d^{(n)}$, making the requirement for moduli
stabilisation even more fundamental for any realistic cosmological setup within
braneworld cosmology.

\section{Acknowledgements}
The authors would like to thank Anne Davis for her supervision and
comments on the manuscript and Andrew Tolley for useful
discussions. SLW is supported by PPARC and CdR by DAMTP.

\appendix
\section{Leading order derivative of the extrinsic curvature}
\label{derivation}
In this appendix, we shall derive an expression for the derivative
of the extrinsic curvature on he branes.
We will not be able to calculate these quantities exactly,
but will be able to obtain a relatively simple expression
for its leading-order contribution.
We will focus on the positive-tension brane first, and our starting point shall be the Taylor series
\be
K\mmix=\sum_{n = 0}^{\infty} \frac{1}{n!}K\nmix{n}, \label{taylor1}
\ee
where we are defining
\[
K^{\mu \, (n)}_{\nu}\equiv\left.\frac{\partial^n}{\partial
y^n}K^{\mu}_{\nu}\right|_{y=0}.
\]
We are interested only in the leading order contribution to $K\nmix{n}$,
\[
K\nmix{n}=\K\nmix{n}\left(1+\mathcal{O}\left(d/L\right)\right),
\]
and the aim of this section will be to establish that
\be
\label{recurrence}
\K\nmix{n}=\OO\K\nmix{n-2},
\ee
where the operator $\OO$ is defined by
\B
\hat O Z^{\mu}_{\nu}&\equiv& \left[
d^{,\mu}Z^{\alpha}_{\ \nu}
+d_{,\nu} Z^{\alpha \mu}-
d^{,\alpha} Z^{\mu}_{\ \nu}
\right]d_{,\alpha}.\nn
\E
This implies that $\K\nmix{n}$ is of the same order as $\K\nmix{n-2}$, and will allow us to produce a simple, albeit formal, expression for this sum, which will be the starting point for writing down the small-$d$ effective theory in the next section.
We will proceed by induction, and throughout make the following assumptions about the order of terms:
\begin{itemize}
\item
$\dm\dn d\sim \dn d\sim d^0$
\item
$\tau\pmix$, $\Dp_\alpha\tau\pmix$ are at worst as divergent as the geometry
\item
$E\pmix\sim d^{-1}$.
\end{itemize}
The assumption on the order of magnitude of the matter terms is
reasonable, since the matter introduced on the brane is expected to
scale as the scale factor for the background and as the curvature
perturbation for general perturbations. Since the curvature
perturbations is in general expected to diverge logarithmically at
the collision, we can hence assume that $\tau\pmix$,
is, at worst, logarithmically divergent in $d$. This implies that
the extrinsic curvature on the brane is itself at worse logarithmically divergent in
$d$.
%
Similarly, we know that $E\pmix\sim d^{-1}$ for the low-energy theory. Although we have argued that
at high energy the moduli space approximation does not give the exact
expression for the Weyl tensor, we have seen that (at least for the
background) the behaviour is the same, differing only in corrections
at higher order in the velocity. In particular $E^{\mu}_{\nu}$ should go as
$d^{-1}$ at high energies as well (we will see later that
this is indeed the case). From (\ref{Euv1}) we have
\B
\label{K'(0)}
K\nmix{1}&=&-dE\mix\left(y=0\right)\ -\left.D^\mu \dn d\right|_{y=0}\nn\\
&&-\frac{\kappa^2 d}{L}\tau\pmix-\frac{\kappa^4}{4}d \tauhp{\mu}{\alpha}\tauhm{\alpha}{\nu}
\E
which, from the above assumptions, gives us
\be
\label{n=1}
\K\nmix{1}\sim d^0.
\ee
For the second derivative of the extrinsic curvature, i.e. for $n=2$,
we need expressions for the derivatives of the Weyl tensor and the Christoffel symbols.
It is straightforward to show that
\be
\Gamma ^{\, \alpha\, \prime}_{\mu \nu}=
D_{\mu}(d\, K^{\alpha}_{\nu})
+D_{\nu}(d\, K^{\alpha}_{\mu}) -D^{\alpha}(d\, K_{\mu \nu}),
\label{gamma '}
\ee
and the derivative of the Weyl tensor is \cite{deRham:2005xv}
\B
E^{\mu\, \prime}_{\ \nu}
&=& \hspace{-5pt}d \Big(2K^{\alpha}_{\nu} E^{\mu}_{\alpha}-\frac{3}{2} K
E^{\mu}_{\nu}-\frac 1 2 K^{\alpha}_{\beta} E^{\beta}_{\alpha}
\delta^{\mu}_{\nu} +C^{\mu}_{\phantom{\mu} \alpha
\nu
\beta}K^{\alpha \beta}\label{E prime}\\
&&2\hat{K}^\alpha_\mu
\hat{K}_{\alpha\beta}
\hat{K}^\alpha_\nu-\frac{7}{6}\hat{K}_{\alpha\beta}\hat{K}^{\alpha\beta}\hat{K}_{\mu\nu}
-\frac{1}{2}q_{\mu\nu}\hat{K}_{\alpha\beta}\hat{K}^\beta_\rho\hat{K}^{\alpha\rho}\Big)\nn\\ && \hspace{-5pt}
-\frac{1}{2 d}D^{\alpha} \left[
d^2 D^{\mu} K_{\alpha \nu}+d^2D_{\nu} K^{\mu}_{\alpha}-2d^2 D_{\alpha}
K^{\mu}_{\nu}
\right],\nn
\E
where $\hat{K}\mix=K\mix-\frac{1}{4}K\delta\mix$.
On the brane, from the Isra\"el matching conditions, the trace of
the extrinsic curvature is
$K\sim\tau$, hence $\hat{K}\mix\sim\tau$ also. So
the cubic terms in $\hat{K}$ will be of higher order than the $KE$ terms, as will the $CK$ term.
The leading terms will, in fact, just be the first three, giving
\be
\label{ep}
{E\mix}'(0)\sim \frac{4d}{L}E\mix(0).
\ee
On the brane,
\B \Dp_\alpha\left(d K\pmix\right)&=&\left(\partial_\alpha d \right) K\pmix + d\Dp_\alpha K\pmix\nn\\
&\sim&\left(\partial_\alpha d \right)K\pmix,
\E
the second term being subdominant from the assumption that
 $\Dp_\alpha K\pmix\propto \Dp_\alpha\tauhp{\mu}{\nu}$ is of higher order than
 $d^{-1}$. The derivative of the Christoffel symbol will similarly
 be of the same order as the extrinsic curvature on the brane:
\be
\label{gammap}
\Gamma ^{\, \alpha\, \prime}_{\mu \nu}(0)\sim
\left(
d_{, \mu}K^{(+)\alpha}_{\gap\nu}
+d_{, \nu}K^{(+)\alpha}_{\gap\mu}
-d^{, \alpha}K^{(+)}_{\mu \nu}
\right).
\ee
Taking the derivative of (\ref{Euv1}) gives
\B {K\mix}''(y)&=&- d {E\mix}'
+2dK^{\mu\beta} D_{\beta} \dn d +
 q^{\mu \beta} \Gamma^{\, \alpha\, \prime}_{\beta \nu}
\partial_{\alpha}d\nn\\&&-d\ \partial_{y}
\left(K^{\mu}_{\alpha}K^{\alpha}_{\nu}\right)
\label{K''},
\E
in the bulk. Evaluated on the brane using (\ref{ep}) and (\ref{gammap}),
the dominant term (of order $d^0$) is the one containing the derivative of the Christoffel symbol
%
\B
\label{order of k2}
K\nmix{2}&\sim& q^{(+)\mu \beta} \Gamma^{\alpha\, \prime}_{\beta
\nu}(0)
\partial_{\alpha}d\sim K.
\E
Since we have shown in (\ref{gammap}) that $\Gamma^{\alpha\, \prime}_{\beta
\nu}(0)\sim d_{, \beta}K^{(+)\alpha}_{\gap\nu}$, on the brane, the second derivative of the extrinsic
curvature is hence of the same order as the extrinsic curvature
itself $K\nmix{2} \sim K\mix$.

Using (\ref{gammap}), we have proved the result for $n=2$:
\B
\K\nmix{2}&=&d_{,\alpha}\left[d^{,\mu} K^{(+)\alpha}_{\gap\nu}+d_{,\nu} K^{(+)\alpha\mu}-d^{,\alpha}K^{(+)}_\mn\right]\nn\\
\label{n=2}&\sim&\OO K\pmix.
\E
The second derivative of the Christoffel symbol follows from (\ref{gamma '}):
\ba
\Gamma ^{\, \alpha\ \prime \prime}_{\mu \nu}(y)&=&
D_\mu\left(d\, K^{\alpha\, \prime}_{\, \nu}\right)+D_\nu\left(d\, K^{\alpha\, \prime}_{\, \mu}\right) -D^\alpha\left(d \, K^{\prime}_{\mu
\nu}\right)\nn \\
&& +d\,\left( \Gamma^{\, \alpha \, \prime}_{\mu \rho}
K^\rho_\nu+\Gamma^{\, \alpha \, \prime}_{\nu
\rho}
K^\rho_\mu
-2 \Gamma^{\,\rho \, \prime}_{\mu \nu} K^\alpha_\rho \right)\label{gamma ''} \\
&& -d\, q^{\alpha \beta}q_{\mu \sigma}\,  \left(
\Gamma^{\, \sigma \, \prime}_{\beta \rho} K^\rho_\nu
-\Gamma^{\, \rho \, \prime}_{\beta \nu} K^\sigma_\rho
\right) \notag\\
&&-d\,\left(q^{\alpha \beta}q_{\mu \sigma}\right)^{\prime}\,
\left(
\Gamma^{\, \sigma}_{\beta \rho} K^\rho_\nu
-\Gamma^{\, \rho }_{\beta \nu} K^\sigma_\rho
\right)\nn\E
where, recall, ${K\mix}'=d\mathcal{L}_n K\mix$ is a tensor, hence the use of covariant derivatives.
These $D(dK')$ terms are of order $d^0$, whilst the others are all of higher order, when evaluated on the brane.
The leading term is
\B
\label{gamma 2}
\Gamma ^{\, \alpha\, \prime \prime}_{\mu \nu}(0)&\sim&
\partial_\mu d\, K^{\alpha\, \prime}_{\, \nu}(0)
+\partial_\nu d\, K^{\alpha\, \prime}_{\, \mu}(0)
-\partial^\alpha d \, K^{\prime}_{\mu \nu}(0)\nn\\
&\sim& d^0.\ea
Substituting (\ref{E prime}) into (\ref{K''})
gives a complicated second-order differential equation for $K\mix$.
Taking repeated $y$-derivatives of this equation would be impractical,
but to start with all we want to do is to work to leading order.
We will first identify which term is dominant, before actually evaluating it.
We therefore drop all indices and numerical factors for the time being, writing $q$ for the metric
(with indices in any position), $K$ for $K\mix$ and $\partial$ for $\dm$.
For example, $q'_{\mn}\propto dK_\mn=d q_{\mu\sigma}K^\sigma_\nu$ and so we would write $q'=dqK$.
The equation for $K''$ can then be written symbolically as
\ba
K^{\prime \prime}&=&
d\, \left(\partial^2 +\Gamma \partial +\partial \Gamma +\Gamma^2 \right)
 q'
+d K^3 +d K \label{K'' symbolic}\\
&& +d K K'+q\,  \partial d \left(\Gamma' +d\, \Gamma K +d\, \partial K
\right)+ \partial^2 d\, q', \notag
\ea
and, from (\ref{order of k2}), we already know that the dominant term is $q\,\partial d\,\Gamma'\sim K$.
We know that $K^{(m)}(0)$ and $\Gamma^{(m)}(0)$ are all of order $d^0$ or $K(0)$ for
$m=0,1,2$.
Recalling that
the extrinsic curvature on the brane is at worse logarithmically divergent in
$d$, terms of the form $d\, K(0)$ will hence be negligible compared
to terms of order $d^0$ (and of course compared to terms of order
$K(0)$). Compared to the $d^{-1}$ divergence, $K(0)$ is hence still
negligible. In what follows, terms of order $d^0$ (such as $K'(0)$, $\Gamma(0)$ and $\Gamma ''(0)$)
and terms of order $K(0)$ (such as $K(0)$, $K''(0)$ and
$\Gamma'(0)$) can hence be treated in a similar way.
Since they are all at worse going as $K(0)$, we shall use in what
follows the notation $K^{(m)}(0)\sim\Gamma^{(m)}(0) \sim K(0)$ for
$m=0,1,2$.
We shall hence take as the inductive hypothesis that this result is true for all $m\le n$. In particular,
\B
\Gamma^{(m)}(0)&\sim&K(0)\quad\forall\ 0\le m\le n\nn\\
q^{(m+1)}(0)&\sim& d\, K(0)\quad\forall\ 0\le m\le n\nn\\
q(0)&\sim& d^0.\nn
\E

Writing $l=n-1$, we have
\B
&& K^{(n+1)}=\partial_y^{(l)}K^{\prime \prime} \label{K(l+2) symbolic}\\
&& \phantom{K^{(n+1)}}=
\partial_y^{(l)}
\Big[
d \left(\partial^2 +\Gamma \partial +\partial \Gamma +\Gamma^2 \right)
 q'
 \nn \\
&&\phantom{K^{(n+1)}\partial_y^{(l)}}
+d K^3+d K +d K K'+ \partial^2 d\, q' \notag \\
&&\phantom{K^{(n+1)}\partial_y^{(l)}}+q\,  \partial d \left(\Gamma' +d\, \Gamma K +d\, \partial K
\right)
\Big]. \notag
\E
Now, evaluating on the brane, we examine the order of each of these terms to find which is the dominant.
For example, remembering that $\partial$ and $\partial_y$ commute, we
have for $l\ge 0$,
\[
\partial_y^{(l)}\left.\left(d\partial^2 q'\right)\right|_{y=0}=d\partial^2\left(q^{(n)}(0)\right)\sim d
K(0)
\]
and similarly
\ba
\left.
\partial_y^{(l)}\Big(d\, \left(\Gamma
\partial +\partial \Gamma +\Gamma^2 \right)
 q'+d\, K^3 +d\, K \Big)\right|_{y=0}&\sim&\hspace{-2pt} d\,K(0)\nn\\
 \left.\partial_y^{(l)}\Big(d\, K K'+q
\partial d\, \left(d \Gamma K
+d \partial K\right)+ \partial^2 d \, q' \Big)\right|_{y=0}& \sim & d\,K(0).\nn
\ea
Finally,
\[
\left. \partial_y^{(l)}\left(q \Gamma' \partial d
\right)\right|_{y=0}= \partial d\, \sum_{m = 0}^{l}\binomial{m}{l}
q^{(l-m)}(0)
\Gamma^{(l+1)}(0),
\]
and the $l=0$ term dominates this last sum, being of order $K(0)$.
Hence the dominant term in the expression for $K\nmix{n+1}$ is,
as in the $K\nmix{2}$ case, the one with the derivative of the Christoffel symbol,
of the same order as $K(0)$.

We have now proved half of the inductive hypothesis, but still need to show that
$\Gamma^{(n+1)}(0)\sim K(0)$. From (\ref{gamma '}), we have
\ba
\Gamma ^{\, \alpha\, (n+1) }_{\mu \nu}(y)
=&&
 \partial_{\mu}(d\, K^{\alpha(n)}_{\nu})
+\partial_{\nu}(d\, K^{\alpha(n)}_{\mu}) -\partial^{\alpha}(d\, K^{(n)}_{\mu \nu})\nn\\
\label{gamma n+1}
&&
+d\, \partial_y^{(n)}
\Big[\Gamma K\ \mathrm{terms}]\\
&& \hspace{-5mm}-\,\sum_{m=0}^{n-1}
\binomial{m}{n-1}
\left(d K^{\rho\, (m)}_{\, \nu}\right)_{,\,
\beta} \partial_y^{(n-m)}\left(q^{\alpha
\beta}q_{\mu \rho}\right).\nn
\ea
By the inductive hypothesis
\[
d\partial_y^{(n)}\left(\Gamma K\right)\sim d\,K(0),\qquad \partial_y^{(n-m)}\left(q^{\alpha\beta}_{\mu\rho}\right)\sim d\,K(0)\ \ n>m
\]
and
\[
\partial_\alpha\left(d
K\nmix{n}\right)=d_{,\alpha}K\nmix{n}+\mathcal{O}(d\,K(0)).
\]
Therefore we can now read off from (\ref{gamma n+1}) the leading order contribution to $\Gamma^{(n)}$,
\ba
\label{leading gamma}
\Gamma ^{\, \alpha\, (n+1)}_{\mu \nu}(0)&=&
d_{,\mu} K^{ \alpha \, (n)}_{\, \nu}
+d_{,\nu} K^{\alpha\, (n)}_{\, \mu}
-d^{,\alpha} K^{(n)}_{\mu \nu}\nn\\&\sim& K(0),
\ea
which agrees with (\ref{gamma 2}) and proves the inductive hypothesis.

We are now finally in a position to calculate $\K\nmix{n}$ for general $n$. We know that the leading contribution obtained from repeatedly differentiating the right-hand side of (\ref{K''}) is
\[
K\nmix{n}\sim q^{\mu \beta} \Gamma^{\, \alpha\,(n-1)}_{\beta \nu}
\partial_{\alpha}d\]
which, from (\ref{leading gamma}), immediately gives (\ref{recurrence}) and the result is proved.

\section{Scalar Perturbations in an FRW background}
\label{appendix scalar perturb}
In this appendix, we shall present some of the details for the
calculation of scalar perturbations of section \ref{section scalar
  perturb}, with the metric perturbation as given in (\ref{ds
  scalar}). We recall that we picked the comoving gauge for which
$\delta d=0$. In that gauge we then have:
\[
\delta\dd = \dot d \Phi
\]
Terms that appear to be sub-dominant will only be dropped at the end.
Using
(\ref{R2}), we get:
\be
\delta R=\frac{\kappa^2}{L}\left(\delta\rho-3 \delta p\right). \label{EqRR}
\ee
Since $a''=0$ for the background (we assume the brane to be empty),
(\ref{EqRR}) implies
\B
\Psi''&+& \frac{a'}{a} \left(2k^2\, E+\Phi'+3
\Psi'\right)\label{RR}\\
&+&\frac{k^2}{3}\left(2 \Psi + 2 E'-\Phi\right)
=\frac{\kappa^2}{6L} a^2\left(\delta\rho-3\delta p\right).\nn
\E
We now perturb (\ref{effectiveeinstein}), writing
\B
z&=&\dd \nn \\
f(z)&=&z\tanh(z/2)\nn\\
g(z)&=&\frac{1}{z}\left(\tanh(z/2)+\tan(z/2)\right)\nn
\E
for simplicity. The $ij$ (with $i\neq j$) component of the Einstein
equation,
to first order in the
perturbations, reduces to:
\be
\Phi-\Psi-2 E'=\left(4 \frac{a'}{a}+2 \frac{d'}{d}\right)E,\label{Gij}
\ee
and the $0i$-component to:
\be
\Psi'=-\left(\frac{a'}{a}+\frac{d'}{2d}\right)\Phi.
\label{G0i}
\ee
So far these equations are equivalent to those one would have obtained
in the low-energy limit. The difference comes from the $00$-component
of the perturbed Einstein equations:
\B
&&\hspace{-10pt}\frac{d'}{d}\Phi'+\frac{a^2}{L d}\left(2f-\dot d f'+\dot d^3
g'\right)\Phi-2k^2\Psi-6\H\Psi'\nn\\
 \label{G00}
&&\hspace{10pt}-4k^2\H E=-\frac{\kappa^2}{6}a
\frac{d'}{d}
\cot \dot d \left(2\delta \rho+3\delta p\right)
\E
and from the equation of motion for $d$:
\B
&&\hspace{-10pt} \Phi'
+\left(8 \frac{d'a'^2}{L}f+
\frac{a}{L}\left(g'\dot{d}^2-4f'\right)\right)\Phi+3\Psi'\label{Eq0}\\
&& \hspace{10pt}
+2k^2 E=-\frac{\kappa^2}{6}a
\left(\cot \dot d \left(2 \delta\rho+3\delta p\right)
-3\coth \dot d\,\delta \rho\right).\nn
\E
Note that one must, at this order, treat $\Phi$,$\Phi'$,$\Psi$ and
$\Psi'$ as four independent variables; differentiation with respect to
conformal time will miss terms arising from higher order in $d$, since
$d$ and $d'$ are of different order. We must then solve the five
equations (\ref{RR}-\ref{Eq0}) simultaneously. Using (\ref{Gij}) to
eliminate $\Phi$ from (\ref{RR}), we obtain
\B
4k^2\left(\frac{d'}{d}-\H\right)E&=&\hspace{-2pt}6\,\Psi''+2 k^2 \Psi
+6\H\left(\Phi'+3\Psi' \right)\label{E}\hspace{15pt}\\
&&\hspace{-5pt}-\frac{\kappa^2}{L}a^2\left(\delta\rho-3\delta p\right)\nn
\E
We may use a combination of (\ref{G00}) and (\ref{Eq0}) to find an
expression for $\Phi'$ in terms of $\Phi, \Psi, \rho$ and $p$ and
hence write $E$ in terms of $\Psi, \Psi', \Psi'', \rho$ and $p$. This
can then be used in (\ref{G00}) to obtain a complicated expression for $\Psi$ in
terms of $\Psi'$,$\Psi''$,$\delta\rho$ and $\delta p$. We then only
keep the leading order in $d$ for each coefficient, resulting in the
much simplified equation
\be
\Psi''+\frac{ d'}{d}\Psi'+k^2\Psi
=\frac{\kappa^2}{6}a^2\frac{\dot{d}}{d}\coth \dot d \
\delta \rho.
\ee

\end{document}